\documentclass[preprint,superscriptaddress,showpacs,preprintnumbers, floatfix]{revtex4}

\usepackage{graphicx}
\usepackage{dcolumn}
\usepackage{bm}
\usepackage{epsfig}

\def\RESULT{0.463~^{+0.073}_{-0.065}~(\mathit{stat}) \pm 0.036~(\mathit{syst})~{\rm ps}}
\def\RESULTCT{139~^{+22}_{-20}~(\mathit{stat}) \pm 11~(\mathit{syst})~{\rm \mu m}}

\begin{document}


\preprint{FERMILAB-PUB-2006/056-E}

\title{\large\bf	Measurement of the \boldmath {$ B_c^+$} Meson Lifetime Using 
\boldmath{ $ B_c^+ \rightarrow J/\psi e^+ \nu_e$} }

\affiliation{Institute of Physics, Academia Sinica, Taipei, Taiwan 11529, Republic of China} 
\affiliation{Argonne National Laboratory, Argonne, Illinois 60439} 
\affiliation{Institut de Fisica d'Altes Energies, Universitat Autonoma de Barcelona, E-08193, Bellaterra (Barcelona), Spain} 
\affiliation{Baylor University, Waco, Texas  76798} 
\affiliation{Istituto Nazionale di Fisica Nucleare, University of Bologna, I-40127 Bologna, Italy} 
\affiliation{Brandeis University, Waltham, Massachusetts 02254} 
\affiliation{University of California, Davis, Davis, California  95616} 
\affiliation{University of California, Los Angeles, Los Angeles, California  90024} 
\affiliation{University of California, San Diego, La Jolla, California  92093} 
\affiliation{University of California, Santa Barbara, Santa Barbara, California 93106} 
\affiliation{Instituto de Fisica de Cantabria, CSIC-University of Cantabria, 39005 Santander, Spain} 
\affiliation{Carnegie Mellon University, Pittsburgh, PA  15213} 
\affiliation{Enrico Fermi Institute, University of Chicago, Chicago, Illinois 60637} 
\affiliation{Joint Institute for Nuclear Research, RU-141980 Dubna, Russia} 
\affiliation{Duke University, Durham, North Carolina  27708} 
\affiliation{Fermi National Accelerator Laboratory, Batavia, Illinois 60510} 
\affiliation{University of Florida, Gainesville, Florida  32611} 
\affiliation{Laboratori Nazionali di Frascati, Istituto Nazionale di Fisica Nucleare, I-00044 Frascati, Italy} 
\affiliation{University of Geneva, CH-1211 Geneva 4, Switzerland} 
\affiliation{Glasgow University, Glasgow G12 8QQ, United Kingdom} 
\affiliation{Harvard University, Cambridge, Massachusetts 02138} 
\affiliation{Division of High Energy Physics, Department of Physics, University of Helsinki and Helsinki Institute of Physics, FIN-00014, Helsinki, Finland} 
\affiliation{University of Illinois, Urbana, Illinois 61801} 
\affiliation{The Johns Hopkins University, Baltimore, Maryland 21218} 
\affiliation{Institut f\"{u}r Experimentelle Kernphysik, Universit\"{a}t Karlsruhe, 76128 Karlsruhe, Germany} 
\affiliation{High Energy Accelerator Research Organization (KEK), Tsukuba, Ibaraki 305, Japan} 
\affiliation{Center for High Energy Physics: Kyungpook National University, Taegu 702-701, Korea; Seoul National University, Seoul 151-742, Korea; and SungKyunKwan University, Suwon 440-746, Korea} 
\affiliation{Ernest Orlando Lawrence Berkeley National Laboratory, Berkeley, California 94720} 
\affiliation{University of Liverpool, Liverpool L69 7ZE, United Kingdom} 
\affiliation{University College London, London WC1E 6BT, United Kingdom} 
\affiliation{Centro de Investigaciones Energeticas Medioambientales y Tecnologicas, E-28040 Madrid, Spain} 
\affiliation{Massachusetts Institute of Technology, Cambridge, Massachusetts  02139} 
\affiliation{Institute of Particle Physics: McGill University, Montr\'{e}al, Canada H3A~2T8; and University of Toronto, Toronto, Canada M5S~1A7} 
\affiliation{University of Michigan, Ann Arbor, Michigan 48109} 
\affiliation{Michigan State University, East Lansing, Michigan  48824} 
\affiliation{Institution for Theoretical and Experimental Physics, ITEP, Moscow 117259, Russia} 
\affiliation{University of New Mexico, Albuquerque, New Mexico 87131} 
\affiliation{Northwestern University, Evanston, Illinois  60208} 
\affiliation{The Ohio State University, Columbus, Ohio  43210} 
\affiliation{Okayama University, Okayama 700-8530, Japan} 
\affiliation{Osaka City University, Osaka 588, Japan} 
\affiliation{University of Oxford, Oxford OX1 3RH, United Kingdom} 
\affiliation{University of Padova, Istituto Nazionale di Fisica Nucleare, Sezione di Padova-Trento, I-35131 Padova, Italy} 
\affiliation{LPNHE, Universite Pierre et Marie Curie/IN2P3-CNRS, UMR7585, Paris, F-75252 France} 
\affiliation{University of Pennsylvania, Philadelphia, Pennsylvania 19104} 
\affiliation{Istituto Nazionale di Fisica Nucleare Pisa, Universities of Pisa, Siena and Scuola Normale Superiore, I-56127 Pisa, Italy} 
\affiliation{University of Pittsburgh, Pittsburgh, Pennsylvania 15260} 
\affiliation{Purdue University, West Lafayette, Indiana 47907} 
\affiliation{University of Rochester, Rochester, New York 14627} 
\affiliation{The Rockefeller University, New York, New York 10021} 
\affiliation{Istituto Nazionale di Fisica Nucleare, Sezione di Roma 1, University of Rome ``La Sapienza," I-00185 Roma, Italy} 
\affiliation{Rutgers University, Piscataway, New Jersey 08855} 
\affiliation{Texas A\&M University, College Station, Texas 77843} 
\affiliation{Istituto Nazionale di Fisica Nucleare, University of Trieste/\ Udine, Italy} 
\affiliation{University of Tsukuba, Tsukuba, Ibaraki 305, Japan} 
\affiliation{Tufts University, Medford, Massachusetts 02155} 
\affiliation{Waseda University, Tokyo 169, Japan} 
\affiliation{Wayne State University, Detroit, Michigan  48201} 
\affiliation{University of Wisconsin, Madison, Wisconsin 53706} 
\affiliation{Yale University, New Haven, Connecticut 06520} 
\author{A.~Abulencia}
\affiliation{University of Illinois, Urbana, Illinois 61801}
\author{D.~Acosta}
\affiliation{University of Florida, Gainesville, Florida  32611}
\author{J.~Adelman}
\affiliation{Enrico Fermi Institute, University of Chicago, Chicago, Illinois 60637}
\author{T.~Affolder}
\affiliation{University of California, Santa Barbara, Santa Barbara, California 93106}
\author{T.~Akimoto}
\affiliation{University of Tsukuba, Tsukuba, Ibaraki 305, Japan}
\author{M.G.~Albrow}
\affiliation{Fermi National Accelerator Laboratory, Batavia, Illinois 60510}
\author{D.~Ambrose}
\affiliation{Fermi National Accelerator Laboratory, Batavia, Illinois 60510}
\author{S.~Amerio}
\affiliation{University of Padova, Istituto Nazionale di Fisica Nucleare, Sezione di Padova-Trento, I-35131 Padova, Italy}
\author{D.~Amidei}
\affiliation{University of Michigan, Ann Arbor, Michigan 48109}
\author{A.~Anastassov}
\affiliation{Rutgers University, Piscataway, New Jersey 08855}
% ...
\collaboration{CDF Collaboration}
\noaffiliation
	


\begin{abstract}

We present a measurement of the $B_c^+$ meson lifetime in the
decay mode 
$B_c^+~\rightarrow  J/\psi e^+ \nu_e$ using the
CDF II detector at the Fermilab Tevatron Collider.  From a 
sample of  about 360~${\rm pb^{-1}}$  of  ${\rm p \bar{p}}$ collisions at  
$\sqrt{s}=1.96~{\rm TeV}$, we reconstruct $J/\psi e^+$ pairs with 
invariant mass in the kinematically allowed range 
$4<M_{J/\psi e}<6$\,GeV/$c^2$.  A fit to the decay-length   
distribution of 238 signal events 
yields a measured $B_c^+$ meson lifetime of  $\RESULT$.

\end{abstract}
\maketitle

The $B_c^+$ meson is the only known meson consisting of two heavy quarks
of different flavor: a charm quark and a bottom antiquark.  It
provides a unique test of heavy-quark dynamics, since the bound state
can be  treated using the same non-relativistic
expansion that successfully describes both  $c \bar{c}$ and 
 $b \bar{b}$ families.  However, unlike 
$c \bar{c}$ and $b \bar{b}$ states, the 
$B_c^+$ meson decays only via weak interactions,  thus having a measurable 
lifetime.  The lifetime of the $B_c^+$ meson  
is expected to be about two to three times 
smaller than the $B^+$ meson lifetime if one assumes three major 
decay subprocesses~\cite{Kiselev,Godfrey}: $\bar{b}$ quark  decay 
with $c$ quark as 
spectator, $c$ quark decay  with $\bar{b}$ quark as 
spectator, and   $\bar{b} c $ annihilation decays. 
In the $B^+$ meson case, the dominant decay subprocess is 
the $\bar{b}$ quark decay with $u$ quark  as spectator.
An early measurement  from CDF~\cite{RunIBc}  found a  $B_c^+$ lifetime  
consistent with predictions.   More  precise measurements  
will  determine the relative importance of the three decay subprocesses and
provide insight into the strong dynamics of heavy quarks.  
Here, we report a new $B_c^+$ meson  lifetime measurement 
using the decay mode $B_c^+ \rightarrow J/\psi e^+ \nu_e$, 
where charge-conjugate modes are implied, 
from a sample of ~360~${\rm pb^{-1}}$  
collected 
during 2002-2004 
with the CDF II detector~\cite{JpsiPRD}  at the Fermilab Tevatron
Collider at a center of mass energy of $1.96~{\rm TeV}$.  
The new measurement is about 2.5 times more precise than the previous one.

The $B_c^+ \rightarrow J/\psi e^+ \nu_e$ reconstruction 
starts with $J/\psi \rightarrow \mu^+ \mu^-$ candidates  
selected based upon a two-muon topology  
by the CDF trigger system~\cite{Trigger}.  
The $J/\psi$ candidates are further purified during 
offline reconstruction  by  vertex constraining 
the $\mu^+ \mu^-$ pairs and  by selecting the pairs with    
with momentum transverse to  the beam line
$p_{T_{J/\psi}}$$>$3\,GeV/$c$.  Then, each $J/\psi$ 
candidate with 
a reconstructed mass  within 50\,MeV/$c^2$ of its nominal value is 
combined with an electron to form a $B_c^+$ candidate.  

Electron identification uses both specific ionization ($\mathit{dE/dx}$) 
information  from the Central Outer Tracker (COT) 
and calorimeter shower information from the Central ElectroMagnetic 
calorimeter (CEM).  The logarithm of the ratio of the
measured $\mathit{dE/dx}$ value from a charged particle to 
that expected for an electron,
$Z_e = \ln(\mathit{dE/dx}) - \ln(\mathit{dE/dx})_{\rm predict}$, is 
compared to its standard deviation $\sigma_{Z_e}$.  
The expected   $\mathit{dE/dx}$  and $\sigma_{Z_e}$ are functions of 
the particle   charge, momentum,  and the  multiplicity of 
associated COT hits.  
Electron candidates are required to have $Z_e/\sigma_{Z_e}>-1.3$ 
to reject hadrons ($\pi/K/p$) 
while  remaining efficient for true electrons.
Samples of electrons, pions, kaons, and protons selected from collision data 
are used to determine  the  $\mathit{dE/dx}$ identification efficiencies 
listed in Table~\ref{eID-eff}.  
These control samples come from photon conversions,   
 $\gamma \rightarrow e^+e^-$, and from  hadron decays,  
$K^0_s \rightarrow \pi^+\pi^-$,  $D^0~\rightarrow~K^-~\pi^+$, 
and $\Lambda^0\rightarrow p\pi^-$.
The hadrons surviving the 
$Z_e/\sigma_{Z_e}$ selection are mainly pions, which 
are rejected using calorimeter shower shape information.

The calorimeter shower shape of a charged particle 
with $p_T$$>$2\,GeV/$c$ 
is obtained by extrapolating its track reconstructed in the COT 
into the calorimeter to match shower clusters there~\cite{RunIBc}.  
The probabilities for a particle to have a shower
shape consistent with being an electron or a hadron are calculated
using the distributions of shower energy and shower cluster profiles
for the electron and hadron samples described above. 
We first define the probability for a charged particle to be an electron 
based on shower shape  by the ratio between its  probability to be 
an electron and the sum of  its probabilities to be an electron or  a hadron. 
We then obtain a cumulative probability distribution 
of the ratio using the  electron sample and  impose a selection on electron
candidates at a 70\% probability  value.  
To calculate the average probability for hadrons to pass this
requirement, as listed in Table~\ref{eID-eff}, the control 
samples of   $\pi/K/p$  particles
are mixed using fractions  predicted by a  {\sc PYTHIA}
Monte Carlo (MC) simulation of $B\rightarrow J/\psi X$ events~\cite{PythiaMC}. 
In addition, electrons found to   originate from photon 
conversion,  $\gamma \rightarrow e^+ e^-$, are removed 
from consideration   as $J/\psi e^+$ candidates~\cite{RunIBc}. 
Overall, the electron identification using 
combined $\mathit{dE/dx}$ and calorimeter 
information has an     efficiency of 60\% for true electrons, while  
hadrons  have a  probability to pass the selection lower than  0.15\%.  

\begingroup
\squeezetable
\begin{table*}[ht]
\caption{Electron identification efficiencies (\%) as functions of particle 
$p_T$ (GeV/$c$) using $\mathit{dE/dx}$ and  
calorimeter ($\mathit{Cal}$) for electron and  hadrons.  
The $\mathit{dE/dx}$  results are averages 
of positively and negatively charged particles.  
The calorimeter results for hadrons ($\mathit{h}$) are the 
weighted averages of  $\pi$, $K$ and $p$.   
The conversion-finding  efficiency 
$\epsilon_{conv}$ is also listed. }
\begin{ruledtabular}
\begin{tabular}{|cccccc|} 
$p_T$ & 2-3 & 3-4& 4-5 & 5-6& $>6$ \\ \hline
$\mathit{dE/dx}$:e&$91.3 \pm 0.1 $ & $91.4\pm 0.2$&$90.7\pm 0.2$&$90.5\pm0.3$  
	          &$89.5\pm 0.2$ \\ 
$\mathit{dE/dx}$:$\pi$&$16.4 \pm 0.1 $ & $23.8\pm 0.1$&$32.0\pm0.1$&$39.0\pm0.2$  
	           &$49.4\pm 0.2$ \\ 
$\mathit{dE/dx}$:$K$&$2.09 \pm 0.09 $ & $2.35\pm 0.07$&$3.29\pm0.09$&$4.4\pm 0.1$ 
	           &$9.5\pm 0.2$ \\ 
$\mathit{dE/dx}$:$p$&$3.38 \pm 0.03 $ & $2.14\pm 0.04$&$2.54\pm0.07$&$3.0\pm 0.1$  
	           &$4.9\pm 0.2$ \\
$\mathit{Cal}$:$e^+$ &$ 68.5\pm 0.3 $ & $68.8\pm 0.5$&$68.9\pm 0.8$&$68.9\pm 1.1$  
	          &$67.6\pm 1.0$ \\ 
$\mathit{Cal}$:$e^-$ &$ 67.9\pm 0.4 $ & $69.5\pm 0.5$&$69.6\pm 0.7$&$68.1\pm 1.2$  
	          &$68.5\pm 0.9$ \\ 
$\mathit{Cal}$:$h^+$ &$ 0.77\pm 0.04 $ & $0.37\pm 0.04$&$0.37\pm0.03$&$0.29\pm0.04$  
	          &$0.13\pm 0.04$ \\ 
$\mathit{Cal}$:$h^-$ &$ 0.64\pm 0.04 $ & $0.37\pm 0.02$&$0.27\pm0.03$&$0.25\pm0.04$ 
	          &$0.21\pm 0.04$ \\
$\epsilon_{conv}$& $ 49.8 \pm 1.4 $ & $55.0\pm2.2$&$56.5\pm 3.3$&$61.5\pm 4.5$ 
	          &$69.2\pm 3.4$ \\
\end{tabular}
\end{ruledtabular}
\label{eID-eff}
\end{table*}
\endgroup

A $B_c^+$ candidate is a $J/\psi e^+$ pair with transverse momentum 
$p_{T_{J/\psi e}}$$>$5\,GeV/$c$  and invariant mass 
4$<$$M_{J/\psi e}$$<$6\,GeV/$c^2$.  The upper bound on  $M_{J/\psi e}$ is  
simply  the  kinematic limit from the mass of  
$B_c^+$ meson~\cite{PDG}. The lower bound  is set higher than the kinematic 
limit of $M_{J/\psi}$  to  reduce the background 
from  $B_c^+$ semileptonic decays  other than the 
exclusive decay $B_c^+ \rightarrow J/\psi e^+ \nu_e$ to 
 a few percent~\cite{Kiselev,RunIBc}. The opening angle between 
the $J/\psi$ and electron momenta in the transverse plane 
must be within $90^\circ$
 to reduce generic $b\bar{b}$ background that produces
a $J/\psi$ and an electron from different $b$ hadrons.  
Finally,  the tracks of the three daughter particles, $\mu^+$, $\mu^-$, and 
$e^+$, are fit to a common vertex and 
the $B_c^+$ decay length in the transverse 
plane, $L_{xy}$, is calculated as the projection of the displacement of 
the $B_c^+$ vertex from the primary vertex onto the momentum of the 
$J/\psi e^+$ system.  
The  primary vertex position is obtained from run-by-run averages
using samples of prompt tracks.

\begin{figure}[!ht]
\centerline{\psfig{figure=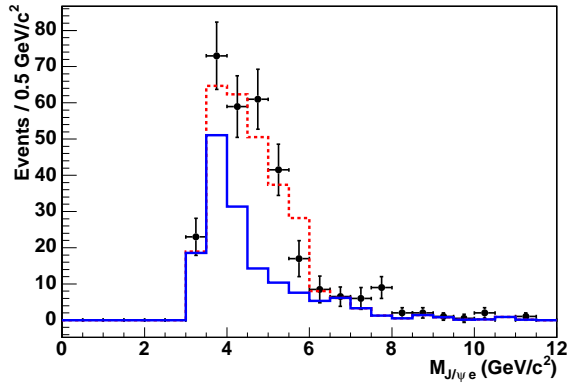,width=0.5\textwidth}}
\caption{$M_{J/\psi e}$ distribution of  $J/\psi e^+$ pairs 
(dots with error bars) with  $L_{xy}/\sigma_{L_{xy}}$$>$3 
together with expected  shape (dashed line) from a sum 
of  a MC signal   and  estimated background (solid lines) 
which is mostly from $b$-hadron decays.  
The false $J/\psi$ is already removed in the distribution.
}
\label{mass_jpsie_ctau}
\end{figure}

Before making a lifetime measurement, we first establish the 
$B_c^+$ signal  in the  $J/\psi e^+$ pairs.
The background pairs  from prompt decays are removed  by 
imposing a selection of  $L_{xy}/\sigma_{L_{xy}}$$>$~3. 
The  $M_{J/\psi e}$  distribution of $J/\psi e^+$ pairs 
with  $L_{xy}/\sigma_{L_{xy}}$$>$~3 is 
shown in Fig.\ref{mass_jpsie_ctau}. Within the 
4$<$$M_{J/\psi e}$$<$6\,GeV/$c^2$
window, 203 candidates are found  with a  total of background 
$88 \pm  14$ as  listed in  Table~\ref{yield-table}. 
 The background is classified into four groups: 
(i) background with a false $J/\psi$, 
background with a correctly identified $J/\psi$ 
but a wrong electron  candidate which can either be (ii)a mis-identified 
hadron (false  electron) or (iii) an electron from 
photon to $e^+ e^-$ conversion,  
and (iv) background from  decays of other B hadrons ($b\bar{b}$). 
The number of false $J/\psi$   
background is estimated using  $\mu^+ \mu^-$ pairs with invariant mass outside 
the 50\,MeV/$c^2$ window.
To estimate the false electron background, we use a sample
of $J/\psi$-track pairs passing the same selection
as the $J/\psi e^+$ pairs, including the $\mathit{dE/dx}$ 
requirement and the requirement to  point to the CEM fiducial
region,  but without any selection based on  calorimeter information. 
 The size of the contribution is estimated from a weighted 
counting of the $J/\psi$-track pairs with the weights taken as 
the averaged probabilities  in Table~\ref{eID-eff}  for hadrons 
to pass the electron selection  using the calorimeter.  
We derive the residual 
conversion electron contribution from the rate of identified photon
conversions together with the conversion-finding 
efficiency listed in Table~\ref{eID-eff}.
The conversion-finding efficiency, defined as the fraction of 
the identified electrons with their conversion partners  in the 
kinematic acceptance of the CDF detector, is estimated from a MC 
simulation. Finally, the contribution 
from decays resulting 
from $b\bar{b}$ production is estimated using a {\sc PYTHIA} MC 
sample with relative rates of flavor creation, flavor excitation, and 
gluon splitting tuned to the Tevatron data~\cite{PythiaMC,RunIBBbar}. 
The number of  $B_c^+$ signal events 
is found  to be 
$115  \pm 16~(\mathit{stat})   \pm 14~(\mathit{syst})$.  
For comparison, there are $2872\pm 59 $  $ B^+ \rightarrow J/\psi K^+$ 
events in the data sample corresponding to the same integrated 
luminosity.   
The selection of $B^+$
is the same as for 
$B_c^+$  except for electron identification on the $K^+$. 
We find the  production rate of $B_c^+$ relative to that of  $B^+$, 
$[\sigma(B_c^+) \cdot
{\cal B} (B_c^+ \rightarrow J/\psi X e^+ \nu_e) ] / [\sigma(B^+) \cdot
{\cal B} (B^+ \rightarrow J/\psi K^+) ]$,  in the kinematic range 
$p_T>4$\,GeV/$c$  and rapidity $|y| <1$~\cite{PDG} to be 
$0.282 \pm 0.038~(\mathit{stat})  \pm 0.035~(\mathit{syst})  \pm 0.065~(\mathit{acceptance})$.  The
first error is statistical; the second covers the systematic uncertainty 
of  $B_c^+$  signal excess counting, and the third pertains to the estimated 
detector acceptance ratio correction  $A_{B^+}/A_{B_c^+} = 4.42 \pm  1.02$
from MC simulation  where the $B_c^+$ $p_T$ spectrum, 
its lifetime values and decay modes   are the major sources of uncertainty.
This new production ratio result agrees  with the earlier CDF 
measurement~\cite{RunIBc}.

\begingroup
\squeezetable
\begin{table}[!ht]
\caption{ Numbers of $J/\psi e^+$ pairs 
and estimated backgrounds.
The error listed  is the sum of 
statistical and systematic errors. }
\begin{ruledtabular}
\begin{tabular}{|lcc|} 
 & All $L_{xy}$   &  $L_{xy}/\sigma_{L_{xy}}>3$ \\ \hline  
 false $J/\psi$ &   $164.0 \pm 9.1$ & $24.5 \pm 3.5$                       \\
 false electron &   $110.2 \pm 19.0$ & $15.4 \pm  2.5 $  \\ 
 conversion electron &  $67.4 \pm 34.8$ & $14.5 \pm 7.8 $  \\
 $b\bar{b}$ &  $63.0 \pm 18.5$ & $33.6 \pm 11.4$  \\ 
prompt decay     &  $141.7 \pm 32.0$  & -		   \\ \hline 
 total background & $545 \pm 55 $ & $88 \pm 14$       \\ 
 observed  $J/\psi e^+$ pairs   & $783$ &$203$   \\ 
\end{tabular}
\end{ruledtabular}
\label{yield-table}
\end{table}
 \endgroup

Having established a clear $B_c^+$ signal in $J/\psi e^+$ 
combinations, we measure the $B_c^+$ meson lifetime in a 
larger sample of 783 
events, selected with the same criteria as above, but without 
the $L_{xy}$ selection.
 We estimate the net signal excess in this
sample to be 238. 
The background sources and their contributions are 
listed in Table~\ref{yield-table}.   
Contributions from false 
electrons, conversion electrons, $b\bar{b}$,  
and false $J/\psi$ are estimated as described earlier.  
The number of  additional prompt-decay events is extracted 
directly from the lifetime fit. 

\begin{figure}[!ht]
\centerline{\psfig{figure=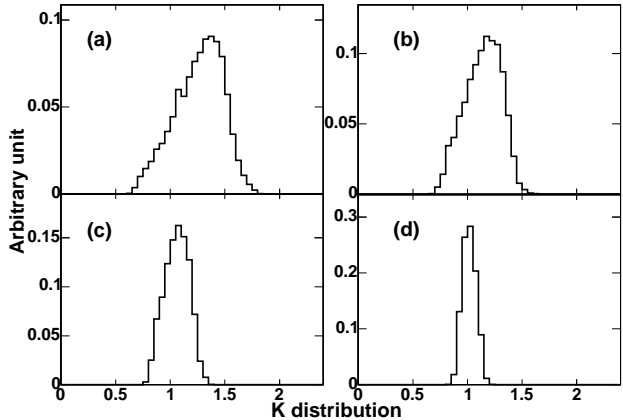,width=0.5\textwidth}}
\caption{Distribution of  $K$ 
for  $M_{J/\psi e}$  within 
(a)~4-4.5, (b)~4.5-5.0, 
(c)~5.0-5.5  and (d)~5.5-6.0\,GeV/$c^2$.  
}
\label{K-factor}
\end{figure}

The Lorentz-invariant proper decay time of a $B_c^+$ event    
is its decay length  $L_{xy}$  with a  Lorentz boost 
 $\beta \gamma = p_{T_{B_c}}/M_{B_c}$ 
in the transverse plane, where 
$M_{B_c}$ is the mass of $B_c^+$ meson and $p_{T_{B_c}}$ its transverse 
momentum.  The $B_c^+$ mass is assumed to have a value of  
6.271\,GeV/$c^2$~\cite{Godfrey}.  
Because of the missing neutrino, we cannot directly 
calculate the boost  factor from the lab system 
to the $B_c^+$ rest frame.  We can however  
calculate the $B_c^+$ lifetime in the  center of mass frame of the 
$J/\psi e^+$ pair, $c t' = L_{xy} M_{J/\psi e}/p_{T_{J/\psi e}}$, 
which provides the best estimator of 
the $B_c^+$  proper decay time in the absence of the neutrino momentum.  
The true proper  $B_c^+$ decay time is given by  
 $c t = L_{xy} p_{T_{B_c}}/M_{B_c} =  K \cdot ct'$,  
where $K$ is the  residual correction 
factor depending on the momenta of the missing neutrino.
In MC simulations,  $K$ can be calculated
$K \equiv   (p_{T_{J/\psi e}} /p_{T_{B_c}})  
\cdot (M_{B_c}/ M_{J/\psi e}) \cdot \cos\alpha$, 
where $\alpha$ is the angle between the vectors of 
$p_{T_{J/\psi e}}$ and $p_{T_{B_c}}$.  
The  $K$ distributions, as shown in  Fig.\ref{K-factor},  always 
include $K$=$1$, and decrease in width as $M_{J/\psi e}$ gets closer to 
the  value of $B_c^+$ mass when the neutrino's momentum is minimal.

\begin{figure}[!ht]
\centerline{\psfig{figure=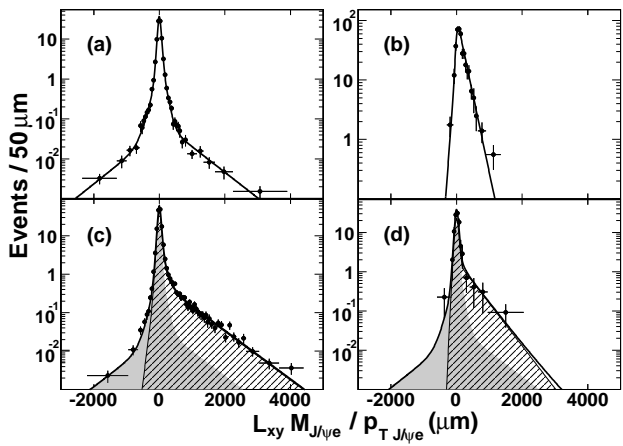,width=0.5\textwidth}}
\caption{ $ct'$  PDF  obtained from fits to background samples 
(points with error bars) for (a)~false $J/\psi$ (solid lines),  
(b)~$b\bar{b}$  (solid lines), (c)~false electron (hatched area) and 
(d)~conversion electron (hatched area).    
In (c) and (d), the solid lines are the sum of false $J/\psi$  
(shaded area)  and false  or conversion electron. 
The false $J/\psi$ fraction and lifetime shape  in (c) and (d)  are 
constrained to that obtained from  sideband events. 
}
\label{lifetime-bkg} 
\end{figure}

 An unbinned maximum-likelihood fit~\cite{RunIBc, JpsiPRD} is used to  
extract the $B_c^+$  meson lifetime.    
In the fit,  $t'$ and its event-by-event error, $\sigma_{t'}  
=  \sigma_{L_{xy}} M_{J/\psi e}/(c p_{T_{J/\psi e}})$,  
are the input variables. The likelihood   function  has 
the form~\cite{Punzi} 
${\cal F}(t',\sigma_{t'}) = 
(1-\sum_1^5 f_{b_i} ){\cal F}_s(t', \sigma_{t'})  
{\cal P}_s (\sigma_{t'}) +
 \sum_1^5 f_{b_i} {\cal F}_{b_i}(t',\sigma_{t'}) {\cal P}_{b_i}(\sigma_{t'})
$,
where  ${\cal F}_s(t',\sigma_{t'})$ is the lifetime 
Probability Density Function (PDF) for  
pure $B_c^+$ signal, $f_{b_i}$ and 
${\cal F}_{b_i}(t',\sigma_{t'})$  are fractions of the five 
background contributions and their lifetime PDFs, and 
${\cal P}_s (\sigma_{t'})$ and  ${\cal P}_{b_i}(\sigma_{t'})$ 
are the PDFs of the $\sigma_{t'}$  for signal and backgrounds.  
The lifetime  PDF for the  $B_c^+$ signal
is an  exponential lifetime distribution   
convoluted with  the $K$ distribution and a 
Gaussian resolution function.   
The prompt background lifetime PDF 
is assumed to have zero lifetime with  a Gaussian resolution function.  
The PDFs for other backgrounds are described by a sum of  
a  Gaussian distribution centered at zero and two pairs of  positive 
and negative  exponential lifetime functions with no $K$ correction applied.   
The initial parameters for these background PDFs are  
obtained   from  fits  to the  background samples as 
shown in Fig.~\ref{lifetime-bkg}. 
The obtained results  are used 
to constrain the corresponding parameters in the final lifetime fit. 
The constraints are imposed by multiplying the likelihood function with 
Gaussian functions  of the appropriate mean and width. 
Similarly, the background fractions  $f_{b_i}$ 
are also constrained 
in the $B_c^+$ meson lifetime fit  
to the estimated values in Table~\ref{yield-table}. 

\begin{figure}
\centerline{\psfig{figure=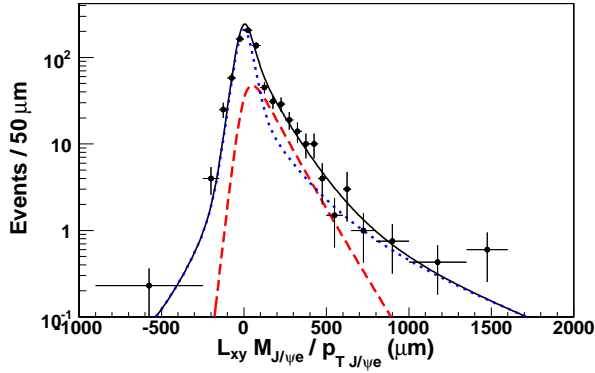,width=0.5\textwidth}}
\caption{$c t'$ distribution from 783 $B_c^+$ candidates.
The points with error bars are data points and the solid  line is the fit 
result.  The dashed line is the $B_c^+$ signal  and 
the dotted line is the background. 
}
\label{lifetime_fit}
\end{figure}

In Fig.\ref{lifetime_fit}, the $c t'$  distribution
from the 783 $B_c^+$ candidates is shown  with the fit result 
superimposed.  
We find  $c\tau_{B_c}= 139^{+22}_{-20}~{\rm \mu m}$. 
The sources of systematic uncertainty 
on the lifetime fit are now considered,  and their 
magnitudes are estimated. 
The effect of the $K$ distribution uncertainty  
on the $B_c^+$ lifetime fit is estimated using alternative 
$K$ distributions obtained from MC 
simulations 
with different $p_T$ spectrum, mass  and 
lifetime values, and $B_c^+$ meson decay modes.
The $p_T$ spectrum is changed from that derived  using a theoretical
calculation~\cite{pTspectrum}  
to that of the inclusive decay $B\rightarrow J/\psi X$~\cite{JpsiPRD}.
The mass and lifetime in the MC calculation are varied in the ranges  
$M_{B_c}=6.2$-$6.4$GeV/$c^2$ and 
$\tau_{B_c}=0.4$-$0.7{\rm ps}$~\cite{Kiselev,Godfrey}.  
The $B_c^+$  decay 
considered in the MC
simulation  
is varied 
from the exclusive $B_c^+ \rightarrow J/\psi e^+ \nu_e$ alone to that of 
an inclusive decay table predicted in Ref.~\cite{Kiselev}.  
We found the change on the $B_c^+$ lifetime fit as
$\Delta c\tau_{B_c} = \pm 2.8~{\rm \mu m}$. 
The uncertainty related to background lifetime shapes 
is  estimated from investigating the $p_T$ dependence of  
the electron identification and  the 
conversion-finding efficiencies, from using false $J/\psi$ events from 
different samples with or without an electron nearby, 
and from changing  fractions of  $b \bar{b}$ events 
originating from the three main  production mechanisms according to a 
study using CDF data~\cite{RunIBBbar}. 
The estimated effect is $\pm 9.2~{\rm \mu m}$.
The uncertainty related to 
the $L_{xy}$ calculation and its error distribution is found 
to be $ \pm 4.7~{\rm \mu m}$ from the uncertainty in the 
silicon detector  alignment,   
by using an  alternative functional form describing  the $L_{xy}$ resolution
 that includes an additional Gaussian  and symmetric exponential tails, 
and by using alternative decay-length error distributions.  
The fitting procedure is also checked using MC 
$c t'$ distributions similar to that of the $B_c^+$ 
candidates and there is no bias found. 
Adding all the estimated systematic  errors in quadrature
we find $c\tau_{B_c}=\RESULTCT$ or $\tau_{B_c}= \RESULT$.


 In conclusion, from an unbinned maximum-likelihood fit to the  
decay-length distribution of 238 signal events  
of $B_c^+ \rightarrow J/\psi e^+  \nu_e$, the $B_c^+$ meson 
lifetime is found to be $\RESULT$ which 
is about one-third  of the $B^+$  meson lifetime of 
$1.671 \pm 0.018~{\rm ps}$~\cite{PDG}.  
This agrees with theoretical models~\cite{Kiselev,Godfrey} 
in which  all the three major decay subprocesses, the two  
spectator processes (${\bar b}$-quark and $c$-quark) and 
the ${\bar b}c$ annihilation, play  important roles in 
the $B_c^+$ decays.

We thank the Fermilab staff and the technical staffs of the participating institutions for their vital contributions. This work was supported by the U.S. Department of Energy and National Science Foundation; the Italian Istituto Nazionale di Fisica Nucleare; the Ministry of Education, Culture, Sports, Science and Technology of Japan; the Natural Sciences and Engineering Research Council of Canada; the National Science Council of the Republic of China; the Swiss National Science Foundation; the A.P. Sloan Foundation; the Bundesministerium f\"ur Bildung und Forschung, Germany; the Korean Science and Engineering Foundation and the Korean Research Foundation; the Particle Physics and Astronomy Research Council and the Royal Society, UK; the Russian Foundation for Basic Research; the Comisi\'on Interministerial de Ciencia y Tecnolog\'{\i}a, Spain; in part by the European Community's Human Potential Programme under contract HPRN-CT-2002-00292; and the Academy of Finland.

\end{document}